\documentclass[twocolumn,nofootinbib,notitlepage,showpacs,prx,amsmath,amstex,amssymb,citeautoscript,longbibliography]{revtex4-1}
\usepackage{amsmath,amssymb,graphicx}
\usepackage{color}
\usepackage{hyperref}
\hypersetup{
    unicode=false,          
    pdftoolbar=false,        
    pdfmenubar=true,        
    pdffitwindow=false,     
    pdfstartview={FitH},    
    pdftitle={Thouless Energy across Many-Body Localization Transition in Floquet Systems},    
    pdfauthor={Michael Sonner, Maksym Serbyn, Zlatko Papic and Dmitry A. Abanin},     
    pdfsubject={},   
    pdfcreator={},   
    pdfproducer={}, 
    pdfkeywords={}, 
    pdfnewwindow=true,      
    colorlinks=true,       
    linkcolor=black,          
    citecolor=blue,        
    filecolor=magenta,      
    urlcolor=blue           
}

\begin{document}
\author{Michael Sonner$^1$, Maksym Serbyn$^2$, Zlatko Papi\'c$^3$ and Dmitry A. Abanin$^1$}

\affiliation{$^1$Department of Theoretical Physics, University of Geneva, 24 quai Ernest-Ansermet, 1211 Geneva, Switzerland}
\affiliation{$^2$IST Austria, Am Campus 1, 3400 Klosterneuburg, Austria}
\affiliation{$^3$School of Physics and Astronomy, University of Leeds, Leeds LS2 9JT, United Kingdom}

\title{Thouless Energy across Many-Body Localization Transition in Floquet Systems}

\begin{abstract}

The notion of Thouless energy plays a central role in the theory of Anderson localization. We investigate and compare the scaling of Thouless energy across the many-body localization (MBL) transition in a Floquet model. We use a combination of methods that are reliable on the ergodic side of the transition (e.g., spectral form factor) and methods that work on the MBL side (e.g. typical matrix elements of local operators) to obtain a complete picture of Thouless energy behavior across the transition. On the ergodic side, Thouless energy tends to a value independent of system size, while at the transition it becomes comparable to the level spacing. Different probes yield consistent estimates of Thouless energy in their overlapping regime of applicability, giving the location of the transition point nearly free of finite-size drift. This work establishes a connection between different definitions of Thouless energy in a many-body setting, and yields new insights into the MBL transition in Floquet systems. 

\end{abstract}

\maketitle

{\it Introduction.--} Out-of-equilibrium properties of disordered
interacting systems have recently attracted much interest. This attention is due to the remarkable fact that such systems may avoid thermalization via the phenomenon of many-body localization (MBL)~\cite{Altman-rev,Huse-rev,AbaninRMP,ALET2018498}. The
characteristic features of MBL, apart from the system's long-term memory of the initial state, include logarithmic spreading of entanglement~\cite{Znidaric08,Moore12} and emergent integrability~\cite{Serbyn13-1,Huse13}, which is stable with respect to finite but
sufficiently weak generic local perturbations. The last property implies that MBL systems represent a paradigm of non-thermal phases of matter, which violate the eigenstate thermalization hypothesis~\cite{DeutschETH, SrednickiETH, RigolNature} at a finite energy density above the ground state. 

In addition to stability with respect to perturbations of the Hamiltonian,  pioneering works~\cite{Lazarides15, Ponte15, Abanin20161, FloquetZhang, SchuchFloquet,lezama_apparent_2019} 
have demonstrated the existence of MBL in the presence of
periodic driving. In these Floquet systems, MBL  allows to avoid unbounded heating, thus enabling the existence
of new non-equilibrium phases of matter, such as time
crystals~\cite{Khemani16,Else16} and anomalous Floquet
insulators~\cite{Po16,Nathan17}. These and related phases are actively
investigated in current experiments with
NV-centers~\cite{Choi16DTC,NVThermalization2018}, cold atoms~\cite{Bordia17}, and trapped
ions~\cite{Zhang16DTC}.

Despite significant recent progress, many open questions remain in the field of MBL, such as the transition between MBL and
delocalized (thermalizing) phase. In studies of localization-delocalization transitions in  {\it single-particle} systems~\cite{kramer_localization_1993,evers_anderson_2008}, a central role is played by the so-called Thouless energy ($E_{\rm Th}$)~\cite{Thouless72}.  Intuitively,  $E_{\rm Th}$ sets the scale at which the system's energy levels develop random-matrix-like correlations. On the one hand, $E_{\rm Th}$ is directly linked to a physical observable -- the system's conductance, while on the other hand, $E_{\rm Th}$ can be defined and practically computed by the response of energy levels to the twisting of boundary conditions. Linking seemingly unrelated characteristics of the system, $E_{\rm Th}$ underlies the celebrated scaling theory of localization~\cite{ScalingTheory}.

The central role of $E_{\rm Th}$ in the understanding of
single-particle localization has motivated its recent extensions to many-body systems. In particular, Ref.~\cite{Serbyn15} introduced a probe based on the behavior of
typical matrix elements of local operators, while
Refs.~\cite{Garcia,Chalker18,KulkarniSFF,sierant_thouless_2020} used spectral properties such
as fluctuations of level number and spectral form factor to map out $E_{\rm Th}$ as a function of disorder strength. Refs. ~\cite{Filippone16,monthus_many-body-localization_2017}, following the original Thouless idea, investigated the
sensitivity of many-body energy levels to boundary conditions. Furthermore, the behavior of the
spectral function was used as a yet another probe of $E_{\rm Th}$~\cite{Serbyn-16}. 
The inverse of $E_{\rm Th}$, the Thouless time, can be understood as a characteristic time scale of the system's dynamics~\cite{schiulaz_thouless_2019}.
It is worth noting that $E_{\rm Th}$ is also one of the central building blocks of
phenomenological renormalization group studies of MBL-thermal
transition~\cite{AltmanRG14,Potter15X}. Thus, several  candidates for the generalization of
$E_{\rm Th}$ to disordered interacting systems have been proposed. However, the comparison of different definitions of $E_{\rm Th}$ is currently missing. Moreover, in Hamiltonian systems, the interpretation of $E_{\rm Th}$ behavior is often complicated by pronounced finite-size effects and non-uniform density of states~\cite{Abanin2019FiniteSize}.

The goal of this paper is to compare the behavior of different notions of $E_{\rm Th}$ in many-body systems. Following Ref.~\cite{FloquetZhang}, we study a many-body
Floquet model without any conservation laws, which reduces finite-size effects compared to the more often studied Hamiltonian models, such as the disordered XXZ spin chain~\cite{PalHuse}. An additional advantage of the Floquet model is that the many-body density of states is uniform, thus removing the need for spectral unfolding~\cite{mehta2004random}. In Hamiltonian models of MBL, on
the other hand, the density of states varies strongly with energy;  in particular, states at the
edge of the spectrum are more localized than those in the center, leading to the many-body mobility edge~\cite{Alet14,Serbyn15,Schiulaz15}.

\begin{figure*}[t]
  \includegraphics[width=\textwidth]{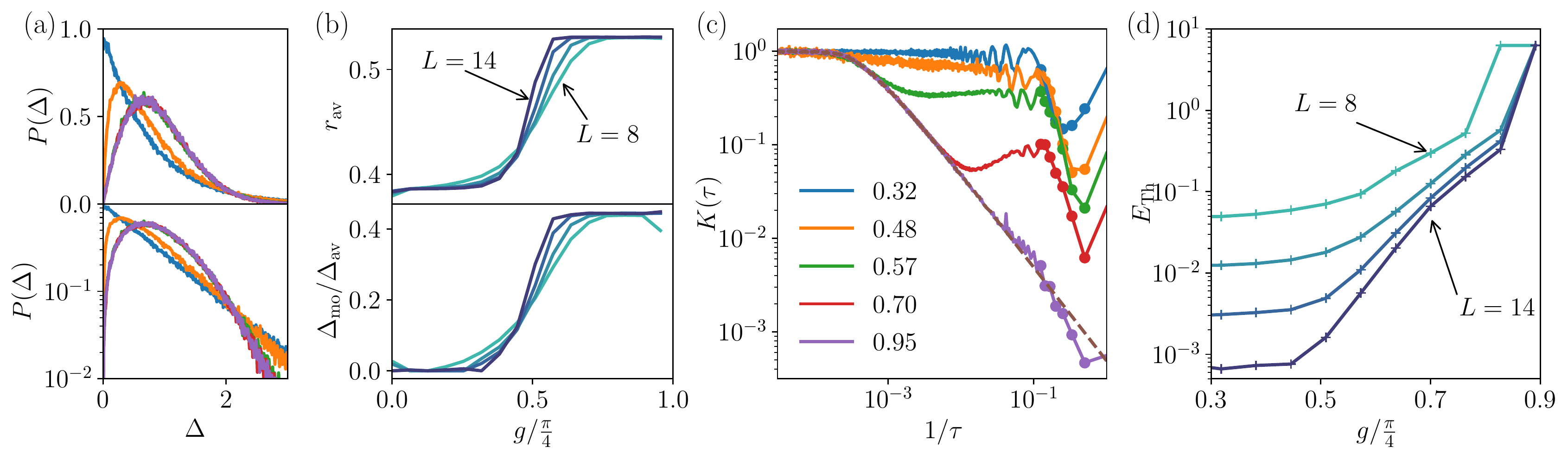}
  \caption{Spectral properties across the MBL transition. (a) Level spacing distribution for different values of disorder $g/\frac{\pi}{4}$, see legend in panel (c). The curves for the three most ergodic values overlap. (b) Average ratio of adjacent level spacings $r_\text{av}$ and the most likely
  level spacing value $\Delta_*$ interpolate between different limiting values
  in the MBL and ergodic phases. Different curves are for system sizes $L=8,\ldots,14$ and their crossing yields
  an estimate for the location of MBL transition $g_* \approx 0.36$.
  (c) Deviation of the spectral form factor $K(\tau)$ from the random matrix
  prediction (dashed line) occurs at progressively larger values of $1/\tau$ as
  disorder strength is decreased ($g$ increases). Data shown for $L=12$ and the
  values of $g/\frac{\pi}{4}$ are specified in the legend. Dots indicate the data
  points obtained with exact disorder averaging using a dual transfer matrix  approach~\cite{Guhr1,BertiniSFF,Sonner20}.
  (d) $E_{\rm Th}$ defined as $1/\tau$ at the point of deviation of $K(\tau)$ from the
  random matrix curve  in panel (c). $E_{\rm Th}$ becomes a constant independent of
  system size on the ergodic side, while in the transition region $E_{\rm Th}$ decreases exponentially with $L$.
  }
  \label{fig:spectrum}
\end{figure*}

Below we demonstrate that a number of different definitions of $E_{\rm Th}$ are
qualitatively consistent  across the MBL transition in the Floquet model defined in Eq.~(\ref{eq:our_model}) below.  At
weak disorder we find that $E_{\rm Th}$ is system-size
independent, consistent with recent analytical results obtained in the large-$N$
limit~\cite{Chalker18}. $E_{\rm Th}$ decreases upon approaching the MBL transition,
becoming smaller than the many-body level spacing. Spectral probes
naturally operate in one of the two regimes: (i) the ergodic regime when $E_{\rm Th}$ is
large, and (ii) the transition regime and MBL phase, where $E_{\rm Th}$ is comparable to or smaller than
level spacing. In contrast, the statistics of matrix elements allows us to access
both regimes. 

 {\it Model.---}We consider the following Floquet model for a periodic chain of $L$ spins-$1/2$,
 defined by the evolution operator over one driving period:
\begin{align}
	\hat{F}=\exp\Big[{-i{\textstyle\sum}_{j} g \sigma_j^x}\Big]\exp\Big[{-i \textstyle\sum_{j}(J \sigma_j^z\sigma_{j+1}^z+ h_j \sigma_j^z)}\Big],
	\label{eq:our_model}
\end{align}
where $\sigma_j^\alpha, \, \alpha=x,y,z$ are Pauli operators, and $h_j \in[0,2\pi]$ are
uniformly distributed random variables. The parameters $g$ and $J$ determine the importance of disorder in
the system. Here, we fix $J=g$ and vary $g$ from $0$, where model is
trivially localized, to $\pi/4$ where the model~(\ref{eq:our_model}) becomes ``perfectly ergodic''~\cite{Guhr1,BertiniSFF,lerose_influence_2020}, in particular it exactly follows certain predictions of random-matrix theory~\cite{BertiniSFF}. 
Furthermore, Refs.~\cite{FloquetZhang,Lezama19} studied a similar Floquet model, with the fields
$h_j$ having both a constant and random components; in contrast, in our model
$h_j$ are fully random. 

In order to study the phase diagram of the model~(\ref{eq:our_model}) and compare
different probes of $E_{\rm Th}$ we use exact diagonalization. We numerically calculate~\cite{SOM} the
quasienergies $\theta_n$ (defined modulo $2\pi$) and eigenvectors $|n\rangle$ of
the Floquet operator, that satisfy $\hat{F}|n\rangle=e^{i\theta_n}|n\rangle$. 
We first extract $E_{\rm Th}$ from the spectral probes which include level statistics and spectral form factor. Afterwards, we proceed with the statistics of matrix elements of local operators which  allows to extract $E_{\rm Th}$ using typical matrix
elements and spectral functions. 

{\it Level statistics.---}Statistics of (quasi)energy levels has long been used as a probe of quantum chaos and integrability in single-particle systems~\cite{Haake2006}, and in recent years it has been fruitfully applied in a many-body setting. It is worth noting that spectral probes may reveal the breakdown of chaos and ergodicity even when conservation laws are not known explicitly.
MBL phase, owing to the emergence of local integrals of motion, exhibits Poisson
level statistics, while the ergodic phase is characterized by level repulsion following Wigner-Dyson, random-matrix level statistics~\cite{PalHuse}. Due to the form of the Floquet operator in Eq.~(\ref{eq:our_model}), the relevant random
matrix theory ensemble is the circular orthogonal ensemble (COE)~\cite{mehta2004random}. These
limiting cases of level statistics for the model~(\ref{eq:our_model}) are demonstrated in
Fig.~\ref{fig:spectrum}(a), which shows the probability distribution of level spacings, $P(\Delta)$, where the level spacing is defined as $\Delta_n =
(\theta_{n+1}-\theta_n)/\delta_\text{av}$, with $\delta_\text{av}=2\pi/2^L$. This definition implies that $\langle \Delta\rangle = 1$, and no unfolding is needed due to constant density of states.

At small values of $g\lesssim 0.3$ corresponding to
strong disorder, the distribution of level spacings $P(\Delta)$ is Poisson, signalling an MBL phase.  At large values of $g$, when the system is deeply in the ergodic phase, $P(\Delta)$
is described by the COE ensemble. At intermediate values of $g$,  level repulsion is still present, as
evidenced by the vanishing of $P(\Delta)$ as $\Delta\to 0$. However, the maximum
$\Delta_*$ of $P(\Delta)$ decreases as $g$ is decreased, compared to the
random-matrix value. This corresponds to the breakdown of the random-matrix
description and weakening of the level repulsion in the critical region, which is also reflected in the softer-than-Gaussian tail of $P(\Delta)$ at large $\Delta$, see bottom panel of 
Fig.~\ref{fig:spectrum}(a).

To estimate the location of MBL-thermal transition from level statistics, we
first use the $r$-parameter~\cite{OganesyanHuse}, defined as  $r_\text{av}= \left\langle {{\rm
min}(\Delta_n,\Delta_{n+1})}/{{\rm max}(\Delta_n,\Delta_{n+1})} \right\rangle$,
where $\langle ..\rangle$ denotes averaging over the spectrum and different
disorder realizations. For the Poisson distribution, this parameter has the value
$r_\text{av}\approx 0.39$, while for the COE-distributed levels it equals
$r_\text{av}\approx 0.54$~\cite{Alessio14}. The behavior of $r_\text{av}$-parameter for different
system sizes is illustrated in Fig.~\ref{fig:spectrum}(b). The curves,
interpolating between the Poisson and COE values at small and large $g$,
respectively, cross at $g_*\approx 0.36$, which we take as the location of the
critical point separating MBL and thermal phases. We note that the drift of the
crossing point with increasing $L$, which is pronounced in Hamiltonian models,
appears to be nearly absent for our model, which we attribute to the absence of
conservation laws~\cite{FloquetZhang}.  

An alternative way of estimating the location of the critical point is provided by studying the most likely value of $\Delta_*$, defined by the maximum of $P(\Delta)$. Parameter $\Delta_*$ is expected to
interpolate between $0$ in the MBL phase and a COE value $\approx 0.45$ in
the ergodic phase. Thus, its finite-size behavior provides a probe of the critical region. Indeed, different curves [Fig.~\ref{fig:spectrum}(b)] cross at
a value consistent with that estimated from $r$-parameter above. 

{\it Spectral form factor.---}Spectral form factor (SFF) probes spectral correlations at energy scales larger than the typical level
spacing~\cite{BrezinSFF}. SFF is given by the  Fourier-transform of the two-level correlation function at ``time"  $\tau>0$,
\begin{equation}\label{eq:SFF}
	K(\tau)=\Big\langle \left| {\rm tr} \left( \hat F^\tau \right) \right |^2 \Big\rangle= \Big\langle \sum_{n,m} e^{i\tau (\theta_n -\theta_m )}\Big\rangle,
\end{equation}
where $\hat F$ is the Floquet operator~(\ref{eq:our_model}),  and $\langle..\rangle$
denotes disorder averaging. As mentioned above, in our Floquet model there is no
need for spectral unfolding, which is necessary for Hamiltonian models. Further,
given that quasienergies are defined modulo $2\pi$, we take $\tau\in Z_+$ taking positive integer values.  Qualitatively, SFF at ``time" $\tau$ probes spectral repulsion, or
its absence at the quasienergy scale $\sim 1/\tau$.

Random-matrix theory \cite{mehta2004random} predicts a linear dependence of SFF
on $\tau$. Ergodic systems are expected to exhibit such linear behavior
$K(\tau)\propto |\tau|$ at energy scales below $E_{\rm Th}$, $1/\tau\lesssim E_{\rm Th}$. Thus, SFF provides a way to extract $E_{\rm Th}$; however, this is only possible in the
ergodic phase, where $E_{\rm Th}$ is large, while in the MBL phase other means should be
sought. We note that at energy scales much smaller than the typical level
spacing, $1/\tau\ll \delta_\text{av}$, the SFF becomes a constant, irrespective
of whether the system is ergodic.

 We computed SFF in two complementary ways: first, by directly evaluating
 Eq.~(\ref{eq:SFF}) from the energy spectrum and averaging over $~100-1000$
 disorder configuration, and, second, using the exactly disorder-averaged
 dual-transfer matrix approach described in Refs.~\cite{Guhr1,Sonner20,BertiniSFF}. The latter method
 is limited to relatively short times $\tau\sim 10$, but provides a useful
 benchmark for confirming that the disorder-averaging in the former method is
 sufficient.

Fig.~\ref{fig:spectrum}(c) shows that in the ergodic phase
($g\gtrsim g_*$), SFF depends linearly on $\tau$ up to $E_{\rm Th}(g,L)$, as predicted by the random-matrix theory. The scaling of $E_{\rm Th}$ for different system
 sizes, shown in Fig.~\ref{fig:spectrum}(d), suggests that in the ergodic
 phase, $E_{\rm Th}(g,L)\to E_{\rm Th}^*(g)$ as $L\to\infty$, that is, Thouless
 energy tends to a constant that does not depend on system size (but depends on $g$). This is consistent with previous results~\cite{Chalker18}, obtained for a model of $q$-state spins with $q\to\infty$ and $d>1$.
In the critical region, we find the scaling $E_{\rm Th}(g,L)\propto
 \delta_\text{av}$, showing that, similar to Hamiltonian systems, at the
 MBL-thermal transition the ratio of $E_{\rm Th}$ and level spacing remains
 approximately constant~\cite{Serbyn15}. Note that the point where $E_\text{Th}$ extracted from SFF exhibits clear exponential scaling with $L$ coincides with the point where $\Delta_*$ remains constant [Fig.~\ref{fig:spectrum}(b), (d)], demonstrating that the two probes yield consistent results. 

\begin{figure}[b]
  \includegraphics[width=\columnwidth]{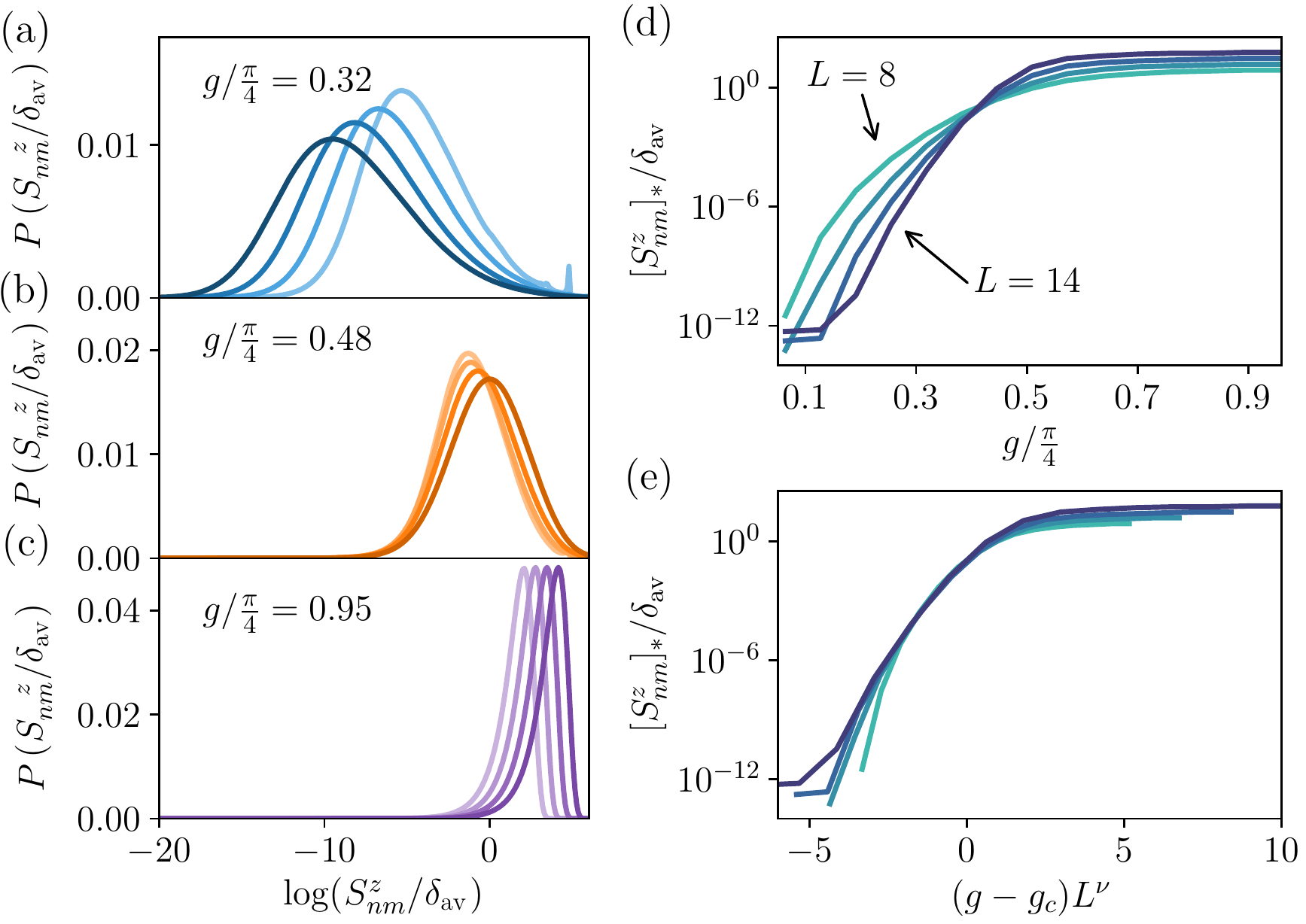}
  \caption{
  (a)-(c) Statistics of ratios between matrix elements and level spacing displays qualitatively different behavior in different phases. 
  Color corresponds to system sizes $L=8,\ldots, 14$ in the order of increasing intensity.
  (a) In the MBL phase the distribution broadens and the typical value decreases exponentially with system size.
  (b) At the transition, the mode moves very little, but the distribution broadens with $L$.
  (c) Deep in the ergodic phase, the evolution of the distribution is consistent with ETH predictions.
  (d)  Mode of the matrix element distribution vs $g$. The crossing point of curves for different $L$ is consistent with the transition estimate from spectral probes.
  (e) Scaling collapse of the data in panel (d) gives the critical exponent $\nu\approx1.2$.
}
    \label{fig:matrix_elements}
\end{figure}

{\it Matrix elements.---}An attractive feature of spectral probes is their
universality: additional conservation laws, irrespective of their precise form,
would lead to Poisson level statistics. However, to describe real-time behavior
of physical observables, and to obtain insights into the structure of conservation
laws in the non-thermalizing phases, it is necessary to study the structure of eigenfunctions and matrix elements of physical operators. Numerically, this comes with an added advantage of utilizing more information per sample. Below we will investigate two ways of extracting $E_{\rm Th}$ from
matrix elements of local operators, which have been employed in Hamiltonian
models of MBL: the ratio between matrix element and level spacing~\cite{Serbyn15} and spectral functions~\cite{LuitzAnomalous,Serbyn-16}.

We first focus on the statistics of matrix elements of a local operator
$\hat{O}$ between the Floquet eigenstates, $O_{nm}=|\langle n | \hat O| m \rangle|$,
and the corresponding ``many-body Thouless parameter" defined here as $\mathcal{G}=
\left[O_{nm}/\delta_{\rm av}\right]_{*}$. The notation $[..]_{*}$ denotes the mode value, this definition is nearly identical to the one that uses an average of the logarithm~\cite{Serbyn15,SOM}. Deep in the ergodic phase,
matrix elements obey the ETH~\cite{DeutschETH,SrednickiETH}, which implies scaling $O_{nm}\propto
\sqrt{\delta_\text{av}} R_{nm}$, where $R_{nm}$ are random,
normal-distributed numbers with a variance of order one. This corresponds to
$\mathcal{G}\propto\delta_{\rm av}^{-1/2}\gg 1$. In the MBL phase, in contrast,
owing to the emergence of local integrals of motion, typical matrix elements
decay much faster than the level spacing $\mathcal{G}(L)\propto 2^{-\kappa L}$
with $\kappa>1$. Further, similar to localized wave functions, the matrix
elements develop broad, log-normal distribution~\cite{evers_anderson_2008}.

We have studied several local operators for the model (\ref{eq:our_model}), and the
resulting distribution for the operator $\hat S^z=\sigma_1^z/2$ is illustrated in
Fig.~\ref{fig:matrix_elements}(a)-(c). As expected, we
observe that in the MBL phase the distribution $P(S^z_{nm}/\delta_{\rm av})$ is
log-normal, and mode of $S^z_{nm}/\delta_{\rm av}$ decreases exponentially with
$L$. On the ergodic side, in contrast, this ratio exponentially increases, as
predicted by ETH, while the distribution remains narrow. In the critical region,
we find that the distribution broadens, while the typical ratio
$P(S^z_{nm}/\delta_{\rm av})$ remains approximately independent of~$L$.

The behavior of $\mathcal{G}$ in Fig.~\ref{fig:matrix_elements}(d) serves as an indicator
of the location of transition, which yields an estimate consistent with spectral
probes. Similar to the finite-size scaling of $r_\text{av}$, we do not find
any significant drift of the crossing point. An interesting question concerns
the relation of $\mathcal G$ and $\Delta_*$. We find that on the
MBL side of the transition, $\mathcal G(L)$ decays much faster compared to the
latter quantity. We attribute this to the fact that $\mathcal G$ probes matrix elements
between eigenstates with quasienergy difference of order one, while
$\Delta_*$ rather probes matrix elements between nearby energy
states, which are enhanced.

In order to obtain the critical exponents, we perform the scaling collapse
of $\mathcal G(L)$ shown in Fig.~\ref{fig:matrix_elements}(e). The value of the critical exponent is $\nu\approx 1.2$, which still violates the Harris criterion~\cite{Harris,Chandran2015}, however this violation is weaker compared to the exponent in Hamiltonian systems~\cite{Alet14}. This, along with the almost absent drift of the crossing point, suggests weaker finite size effects due to the absence of conserved quantities.

{\it Spectral function.---}Spectral function (SF) quantifies the energy structure of a matrix element which can be experimentally probed in absorption spectroscopy. SF of an operator $\hat{O}$ is defined as
\begin{equation} \label{eq:spectral_function}
  f^2(\omega) =2^{-L} \sum_{n,m} |\langle n |
  \hat{O} | m \rangle|^2  \delta(\omega-\theta_n+\theta_m),
\end{equation}
where the sum runs over all eigenstates, corresponding to the infinite temperature
ensemble. This SF is a Fourier transform of the infinite-temperature real-time
correlation function $\langle \hat O(t) \hat O(0) \rangle_{T=\infty}$, thus, it
contains information about the system's dynamical time scales. In particular, in the
ergodic phase, the spectral function has a plateau for $\omega\lesssim E_{\rm
Th}$, since the Thouless time is a scale at which the excitations have propagated
through the system and the dynamics have saturated.

\begin{figure}[t]
  \includegraphics[width=\columnwidth]{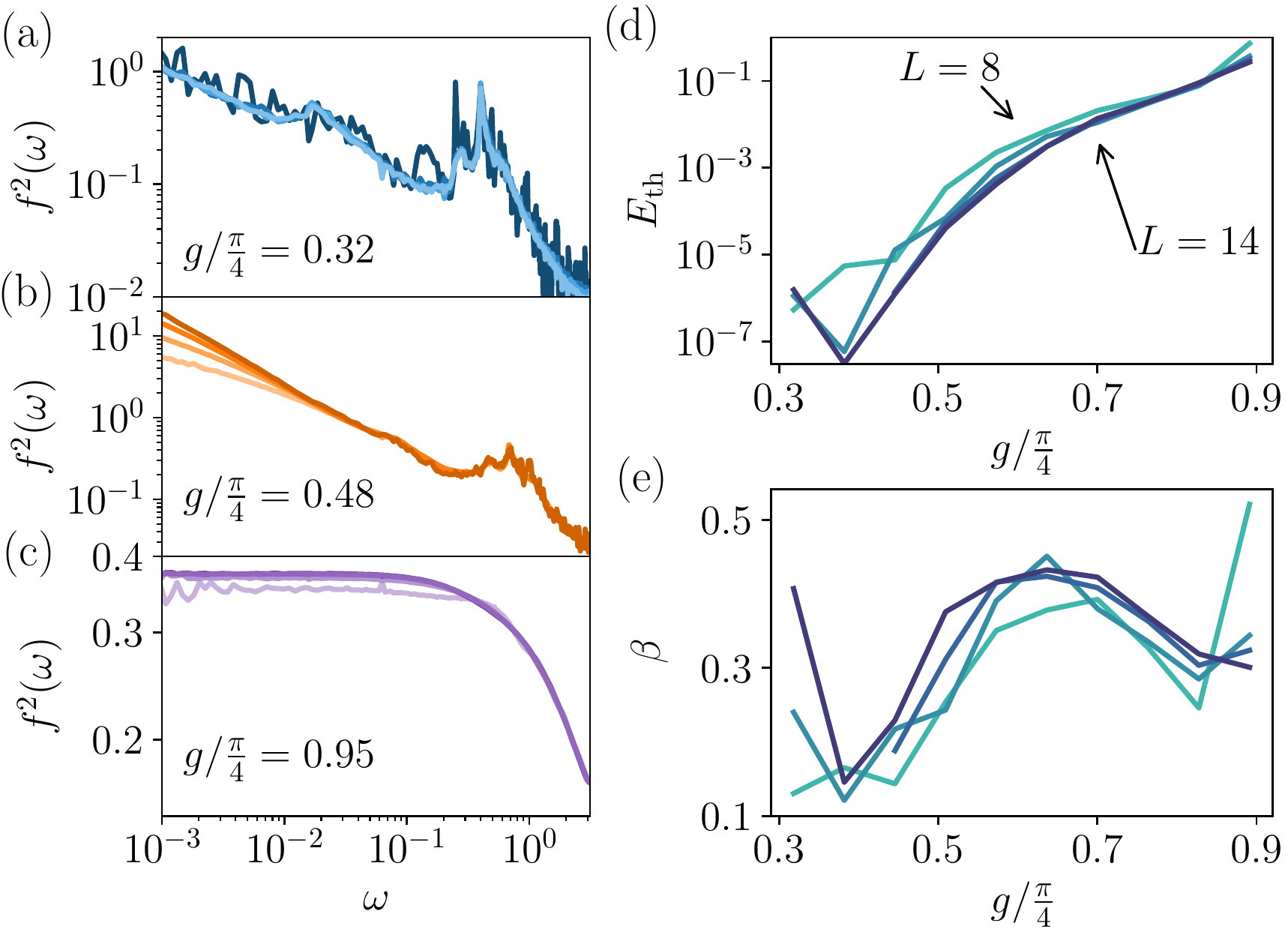}
  \caption{Spectral function (\ref{eq:spectral_function}) for different system sizes in the MBL phase (a), at
  the transition (b), and in the fully ergodic phase (c). For energies larger
  than $E_{\rm Th}$ the spectral function does not depend on system size. Since $E_{\rm Th}$ is very
  small on the MBL side and constant in the ergodic phase, the spectral function
  collapses in the full range of considered frequencies. (d) Scaling of 
  $E_{\rm Th}$ extracted from the kww fit to the spectral function~\cite{SOM}. (e)
  Exponent of the asymptotic power law of the kww fit~\cite{kww}.}
  \label{fig:spectral_function}
\end{figure}

The evolution of disorder-averaged spectral function of $\hat S^z$ operator,
$f^2 (\omega)$, across the MBL-thermal transition is illustrated in
Fig.~\ref{fig:spectral_function}(a)-(c). Deep in the MBL phase,
spectral function develops a delta-function peak at $\omega=0$~\cite{SOM}, while its behavior at $\omega>0$ approximately follows power-law
behavior, also observed in Hamiltonian MBL systems~\cite{Serbyn-16}. In the
critical region, a power-law with a larger exponent gradually
develops, and spectral function does not exhibit a visible plateau, consistent
with $E_{\rm Th}$ becoming of order of the level spacing. Deep in the ergodic phase (c),
in contrast, the spectral function is nearly independent of system size,
showing a wide plateau, which yields an estimate of $E_{\rm Th}$ consistent with spectral
probes. 

To quantify $E_{\rm Th}$ from the spectral function we fit $f^2 (\omega)$
using the Fourier transform of a stretched exponential (so-called kww function~\cite{kww}). This is consistent with recent work~\cite{Lezama19} that reported the stretched exponential decay of
real time correlation functions. The fitting procedure~\cite{SOM} results in $E_{\rm Th}$ shown in Fig.~\ref{fig:spectral_function}(d). The exponent $\beta$, which controls the asymptotic power-law behavior of the spectral function is illustrated in Fig.~\ref{fig:spectral_function}(e). The spectral function $f^2(\omega)$ behaves as $1/\omega^{1+\beta}$ for $\omega\gg E_{\rm Th}$. We note that $\beta$ exhibits non-monotonic behavior with maximum around the transition point. In addition, in vicinity of transition, we observe a collapse of spectral function $f^2(\omega)/2^L$ plotted as a function of $\omega/\delta_\text{av}$ for different system sizes, see~\cite{SOM}. 

{\it Discussion.---} We have studied and compared the behavior of $E_{\rm Th}$ across the Floquet many-body localization
transition obtained using various probes based on spectral properties and matrix elements of local
observables. Among the considered probes, SFF and spectral
function work well on the ergodic side yielding consistent values of $E_{\rm Th}$ that saturates at a value independent of system size. This is in contrast to the Hamiltonian models, where $E_{\rm Th}(L)\propto L^{-1/\gamma}$ exhibits subdiffusive scaling with $L$ with a disorder-dependent exponent~\cite{Agarwal_2017}. In the critical region, $E_{\rm Th}$ becomes of the order of many-body level spacing. In the critical region the
extraction of $E_{\rm Th}$ from the SFF and spectral function becomes unreliable,
and we obtain $E_{\rm Th}$ from the most probable value of the level spacing and many-body
Thouless parameter.  

In addition to scrutinizing different notions of $E_{\rm Th}$, our work provides new
insights into the MBL transition in the Floquet model. The absence of any conserved quantities in the considered model reduces the
finite size effects, which is manifested in the larger value of the critical exponent
and weaker drift of the critical point, in agreement with earlier numerical studies of a different model using a different set of probes~\cite{FloquetZhang}. 
Moreover, we find the properties of spectral functions to be consistent with the stretched
exponential relaxation of real-time correlation functions~\cite{Lezama19}.

Several open questions remain for future work. In particular, in light of an apparent absence of subdiffusion in the ergodic phase in Floquet models, it would be interesting to investigate the rare-region effects on the spreading of entanglement, correlation functions decay, and implications for the nature of the transition into the MBL phase. Furthermore, the link between different definitions of  $E_{\rm Th}$ established here may serve as a foundation for developing a scaling theory of the Floquet-MBL transition. 

{\it Note added.---} While this paper was being finalized, Ref.~\cite{ChalkerMBL20} appeared which investigates SFF in a Floquet model of MBL using dual disorder-averaged transfer matrix approach also employed above.

{\it Acknowledgements.---}
This work was supported by the Swiss National Science Foundation (Mi.So.\ and D.A.), and by the European Research Council (ERC) under the European Union's Horizon 2020 research and innovation programme (Ma.Se., grant agreement No.\ 850899, and D.A., grant agreement No.\ 864597).  Z.P.\ acknowledges support by EPSRC grant EP/R020612/1  and by the Leverhulme Trust Research Leadership Award RL-2019-015.
Statement of compliance with EPSRC policy framework on research data: This publication is theoretical work that does not require supporting research data. The computations were performed on the Baobab cluster of the University of Geneva.

\end{document}